\input harvmac
\input psfig
\newcount\figno
\figno=0
\def\fig#1#2#3{
\par\begingroup\parindent=0pt\leftskip=1cm\rightskip=1cm\parindent=0pt
\global\advance\figno by 1
\midinsert
\epsfxsize=#3
\centerline{\epsfbox{#2}}
\vskip 12pt
{\bf Fig. \the\figno:} #1\par
\endinsert\endgroup\par
}
\def\figlabel#1{\xdef#1{\the\figno}}
\def\U{{\bf U}}
\def\encadremath#1{\vbox{\hrule\hbox{\vrule\kern8pt\vbox{\kern8pt
\hbox{$\displaystyle #1$}\kern8pt}
\kern8pt\vrule}\hrule}}
\def\underarrow#1{\vbox{\ialign{##\crcr$\hfil\displaystyle
 {#1}\hfil$\crcr\noalign{\kern1pt\nointerlineskip}$\longrightarrow$\crcr}}}
%
\overfullrule=0pt

%

\def\bar{\overline}
\def\Z{{\bf Z}}
\def\T{{\bf T}}
\def\S{{\bf S}}
\def\R{{\bf R}}

\font\zfont = cmss10 
\font\litfont = cmr6

\def\bigone{\hbox{1\kern -.23em {\rm l}}}
\def\ZZ{\hbox{\zfont Z\kern-.4emZ}}
\def\half{{\litfont {1 \over 2}}}

\Title{hep-th/9903005, IASSNS-HEP-99-20}
{\vbox{\centerline{Supersymmetric Index}
\bigskip
\centerline{Of Three-Dimensional Gauge Theory}}}
\smallskip
\centerline{Edward Witten}
\smallskip
\centerline{\it School of Natural Sciences, Institute for Advanced Study}
\centerline{\it Olden Lane, Princeton, NJ 08540, USA}\bigskip

\medskip

\noindent

In $N=1$ super Yang-Mills theory in three spacetime dimensions,
with a simple gauge group $G$ and a Chern-Simons interaction
of level $k$, the supersymmetric index $\Tr\,(-1)^F$ can be computed
by making a relation to a pure Chern-Simons theory or 
microscopically by an explicit Born-Oppenheimer calculation on
a two-torus.  The result shows that supersymmetry is unbroken
if $|k|\geq h/2$ (with $h$ the dual Coxeter number of $G$) and suggests
that dynamical supersymmetry breaking occurs for $|k|<h/2$.
The theories with large $|k|$ are massive gauge theories whose
universality class is not fully described by the standard criteria.


\vskip 2cm
\noindent To appear in the Yuri Golfand memorial volume.

\Date{February, 1999}
\newsec{Introduction}

If a $d+1$-dimensional supersymmetric quantum field theory is
quantized on $\T^d\times \R$ (with $\T^d$ understood as space and
$\R$ parametrizing the time), the spectrum is often discrete.
If so, one can define a supersymmetric index $\Tr\,(-1)^F$,
the number of zero energy states that are bosonic minus the number
that are fermionic.  The index is invariant under smooth variations
of parameters (such as masses, couplings, and the flat metric on $\T^d$)
that can be varied while preserving supersymmetry.  For this reason,
it often can be computed even in strongly coupled theories \ref\witten{E.
Witten, ``Constraints On Supersymmetry Breaking,'' Nucl. Phys. {\bf B202}
(1982) 253.}.

When $\Tr\,(-1)^F$ is nonzero, there are supersymmetric states for
any volume of $\T^d$, and hence the ground state energy is zero regardless
of the volume.  When one has a reasonable control on the behavior of the
theory for large field strengths (to avoid for example the possibility
that a supersymmetric state goes off to infinity as the volume goes
to infinity), it follows that the ground state energy is zero and
supersymmetry is unbroken in the infinite volume limit.  Conversely,
if $\Tr\,(-1)^F=0$, this gives a hint that supersymmetry might be
spontaneously broken in the quantum theory, even if it appears to
be unbroken classically.

There are interesting examples of theories (e.g., nonlinear sigma
models in two dimensions, and pure $N=1$ supersymmetric Yang-Mills
theory in four dimensions) in which a nonzero value of $\Tr\,(-1)^F$
has been used to show that supersymmetry remains unbroken even for
strong coupling. But  there are in practice very few instances in which
vanishing of this index has served as a clue to spontaneous supersymmetry
breaking.  One reason for this is that many interesting supersymmetric
theories have a continuous spectrum if compactified on a torus, making
 $\Tr\,(-1)^F$  difficult to define, or have a nonzero value
of $\Tr\,(-1)^F$, so that supersymmetry cannot be broken.
In other examples, $\Tr\,(-1)^F$ is defined and
equals zero, but does not give a useful hint of
supersymmetry breaking because this phenomenon is either obvious classically
or is obstructed by the existence, classically, of a mass gap, or for
other reasons.

The present paper is devoted to 
a case in which the index
does seem to give a clue about when supersymmetry is dynamically broken.
This example is the pure $N=1$ supersymmetric gauge theory in three spacetime
dimensions, with simple compact gauge group $G$.  The theory can be
described in terms of a gauge field $A$ and a gluino field $\lambda$
(a Majorana fermion in the adjoint representation). We include a Chern-Simons
interaction, so the Lagrangian with Euclidean signature reads
\eqn\olpo{L={1\over 4e^2}\int d^3x \Tr\left(F_{IJ}F^{IJ}+\bar\lambda
i\Gamma\cdot D\lambda\right) -{ik\over 4\pi}\int \Tr\left(A\wedge
d A +{2\over 3}A\wedge A\wedge A +\bar\lambda\lambda\right).}
The parameter $k$ is quantized topologically 
\ref\templeton{S. Deser, R. Jackiw, and
S. Templeton, Ann. Phy. (N.Y.) {\bf 140} (1982) 372.}. 
If $h$ denotes the dual Coxeter number of $G$, then the
quantization condition is actually that $k-h/2$ should be
an integer, as  was pointed out in  \nref\who{H.-C. Kao,
Kimyeong Lee, and Taelin Lee, ``The Chern-Simons Coefficient In Supersymmetric
Yang-Mills Chern-Simons Theories,'' hep-th/9506170.}%
\nref\kog{G. Amelino-Camelia, I. I. Kogan, and R. J. Szabo, ``Conformal
Dimensions From Topologically Massive Quantum Field Theory,'' Nucl. Phys.
{\bf B480} (1996) 413, hep-th/9607037.}%
\refs{\who,\kog},
\nref\alv{L. Alvarez-Gaum\'e and E. Witten, ``Gravitational Anomalies,''
Nucl. Phys. {\bf B234} (1983) 269.}%
\nref\redlich{N. Redlich, ``Parity Violation And Gauge Noninvariance
Of The Effective Gauge Field Action In Three-Dimensions,'' Phys. Rev.
{\bf D29} (1984) 2366.}%
using a mechanism of \refs{\alv,\redlich}.  The situation
will be reviewed in section 2.

Let $I(k)$ denote the supersymmetric index as a function of $k$.
We will show that $I(k)\not= 0$ for $|k|\geq h/2$, but $I(k)=0$ for
$|k|<h/2$.  
For example, for $G=SU(n)$, we have $h=n$, and 
\eqn\gobbo{I(k)={1\over (n-1)!}\prod_{j=-n/2+1}^{n/2-1}(k-j).}
So $I(k)$ vanishes precisely if $|k|<n/2=h/2$.  

From this it follows that
supersymmetry is unbroken quantum mechanically
for $|k|\geq h/2$.  But we conjecture that in the ``gap,''
$|k|<h/2$, supersymmetry is spontaneously broken.
For this we offer two bits of evidence beyond the vanishing of the index.
One  is that an attempt to disprove the hypothesis of spontaneous
supersymmetry breaking for $|k|<h/2$ by considering an $SU(n)/\Z_n$ theory
(instead of $SU(n)$) fails in a subtle and interesting way.  
The second is that, as we will see, if the theory is formulated on a two-torus
of finite volume, spontaneous breaking of supersymmetry occurs.
Of course, these considerations do not add up to a proof, but
they are rather suggestive.

The paper is organized as follows.
In section 2, we compute the index for sufficiently large $k$
by using low energy effective field theory and the relation
\ref\wittenjones{E. Witten, ``Quantum Field Theory And The Jones
Polynomial,'' Commun. Math. Phys. {\bf 121} (1989) 351.} 
of Chern-Simons gauge theory
to two-dimensional conformal field theory.  In the process, we also
review the anomaly that sometimes shifts $k$ to half-integer values,
and we explain the failure of a plausible attempt to disprove the
hypothesis of symmetry breaking in the gap via $SU(n)/\Z_n$ gauge theory.
In section 3, we make a more precise microscopic computation of the 
index, and show that for finite volume symmetry breaking does
occur in the gap.    Finally, in section 4, we consider
three-dimensional Chern-Simons theories in the light of the familiar
classification \ref\thooft{G. 't Hooft, ``On The Phase Transition Towards
Permanent Quark Confinement,'' Nucl. Phys. {\bf B138} (1978) 1,
``A Property Of Electric And Magnetic Flux In Non-Abelian Gauge
Theory,'' Nucl. Phys. {\bf B153} (1979) 141, ``Topology Of The
Gauge Condition And New Confinement Phases In
Non-Abelian Gauge Theories,'' Nucl. Phys. {\bf B205} (1982) 1.} 
of massive phases of gauge
theories, and show that such massive phases are not fully classified
by the usual criteria.  This is true even in four dimensions, 
but the full classification
of massive phases is particularly rich in three dimensions.   

For other recent results on dynamics of supersymmetric
Chern-Simons theories in three dimensions, see 
\ref\kap{M. Strassler and A. Kapustin,
``On Mirror Symmetry In Three-Dimensional Abelian Gauge Theories,''
hep-th/9902033.}.

\newsec{Computation Via Low Energy Effective Field Theory}

The index can be computed very quickly if $k$ is sufficiently large.
At the classical level, the theory has a mass gap for $k\not= 0$
\templeton.  The mass is of order $e^2|k|$,
which if $|k|>>1$ is much greater than the scale 
$e^2$ set by the gauge couplings.  So for $|k|>>1$, the classical
computation is reliable, the theory has a mass gap, 
and in particular
(as there is no Goldstone fermion) supersymmetry is unbroken.

Moreover, we can compute the index for sufficiently large $|k|$ using
low energy effective field theory.  
For large enough $|k|$, the mass gap implies
that the fermions can be integrated out to give
a low energy effective action that is still local.
Integrating out the fermions 
gives a shift in the effective value of $k$.  The shift can be computed
exactly at the one-loop level.\foot{There are many ways
to prove this.  For example,  the $s$-loop effective action
for $s>1$ is the integral of a gauge-invariant local density, which the
Chern-Simons functional is not, so a renormalization of the effective
value of $k$ can only come at one loop.  Alternatively, an $s$-loop
diagram is proportional to $e^{2s-2}$, and so can only renormalize
an integer $k$ if $s=1$.}
In fact, integrating out the fermions 
shifts the effective value of $k$ in the low energy effective field
theory to 
\eqn\olop{k'=k-{h\over 2}\,{\rm sgn}(k),}
 where ${\rm sgn}(k)$ is the sign
of $k$ \who.  
(The shift in $k$ is proportional to the sign of $k$, because this
sign determines the sign of the fermion mass term.)
So for example if $k$ is positive, as we assume until
further notice, then $k'=k-h/2$.
For the low energy theory to make sense, $k'$ must be an integer,
and hence $k$ must be congruent to $h/2$ modulo $\Z$.  So if
$h$ is odd, then $k$ is half-integral, rather than integral \who.
For example, for $SU(n)$, $h=n$ and $k$ is half-integral if $n$ is odd.

Since the factor of $h/2$ will be important in this paper, we pause
to comment on how it emerges from Feynman diagrams.  The basic
parity anomaly \refs{\alv,\redlich} is the assertion that for an
$SU(2)$ gauge theory with Majorana fermions consisting of two copies
of the two-dimensional representation, the one-loop shift in $k$ is
$1/2$.  (We must take two copies of the ${\bf 2}$ of $SU(2)$, not one,
because the ${\bf 2}$ is a pseudoreal representation, but Majorana fermions
are real.)  For any other representation, the one-loop shift is scaled
up in proportion to the trace of the
quadratic Casimir for that representation.  For three Majorana fermions
in the adjoint representation of $SU(2)$, the trace of the quadratic
Casimir is twice as big as for two ${\bf 2}$'s, so the shift in $k$ is
$1$, which we write as $h/2$, with $h=2$ for $SU(2)$.  The result
$h/2$ is universal, since $h$ is
the group theory factor in the one-loop diagram
for any group.  This argument does not explain the minus sign in the
formula $k'=k-h/2$, which depends on some care with orientations.
  This sign can be seen in Feynman diagrams \who, and also
has a topological meaning that we will see in section 3.

Now, for very large $k$, though the theory has a mass gap, it is not
completely trivial at low energies.  Rather, there is a nontrivial
dynamics of zero energy states governed by the Chern-Simons theory
at level $k'$.  
At low energies, we can ignore the Maxwell-Yang-Mills term in the action,
and approximate the theory by a ``pure Chern-Simons'' theory, with
the Chern-Simons action only.  This is a topological field theory
and in fact is a particularly interesting one.
In general, if the pure Chern-Simons theory at level
$k'$ is formulated on a Riemann surface $\Sigma$ of genus $g$, then
\wittenjones\ the number of zero energy states equals the number of
conformal blocks of the WZW model of $G$ at level $k'$.  Moreover,
these states are all bosonic.\foot{Or they are all fermionic.
In finite volume, there is a potentially arbitrary sign choice
in the definition of the operator $(-1)^F$, as we will see in more
detail in section 3.}
   For our present
application, $\Sigma =\T^2$ and the genus is 1.  In this case, the
number $J(k')$ of conformal blocks is equal to the number of representations
of the affine Lie algebra $\widehat G$ at level $k'$.  This number is
positive for all $k'\geq 0$, and for large $k'$ is of order $(k')^r$,
with $r$ the rank of $G$.  For more detail on the canonical
\nref\elitz{S. Elitzur, G. Moore, A. Schwimmer, and N. Seiberg,
``Remarks On The Canonical Quantization Of The Chern-Simons-Witten
Theory,'' Nucl. Phys. {\bf B326} (1989) 108.}
\nref\axel{S. Axelrod, S. Della Pietra, and E. Witten, ``Geometric Quantization
Of Chern-Simons Gauge Theory,'' J. Diff. Geom. {\bf 33} (1991) 787.}
\nref\cresci{M. Crescimanno and S. A. Hotes, ``Monopoles, Modular
Invariance, and Chern-Simons Field Theory,'' Nucl. Phys. {\bf B372}
(1992) 683.}
quantization of the Chern-Simons theory, see \refs{\elitz - \cresci}.

\nref\pisarski{R. D. Pisarski and S. Rao, ``Topologically Massive
Chromodynamics In The Perturbative Regime,'' Phys. Rev. {\bf D32} (185) 2081.}
\nref\manny{W. Chen, G. W. Semenoff, and Y.-S. Wu,``Finite
Renormalization Of Chern-Simons Gauge Theory,'' Mod. Phys. Lett.
{\bf A5} (1990) 1833; L. Alvarez-Gaum\'e, J. M. F. Labastida and
A. V. Ramallo, ``A Note On Perturbative Chern-Simons
Theory,'' Nucl. Phys. {\bf B334} (1990) 103; M. Asorey and F. Falceto,
``Chern-Simons Theory And Geometric Regularization,''
Phys. Lett. {\bf B241} (1990) 31; C. P. Martin, 
``Dimensional Regularization Of Chern-Simons Field Theory,''
Phys. Lett.
{\bf B241} (1990) 513; M. A. Shifman, ``Four-Dimensional Aspect Of
The Perturbative Renormalization In Three-Dimensional Chern-Simons
Theory,''
Nucl. Phys. {\bf B352} (1991) 87; I. I. Kogan, ``Topologically Massive Gauge
Theories: Who Needs Them And Why?'' Comments Nucl. Part. Phys. {\bf A19}
(1990) 305, ``Quantum Mechanics On The Moduli Space From The Quantum
Geometrodynamics Of The Open Topological Membrane,'' Phys. Lett. {\bf B256} 
(1991) 369.}
The following paragraph is aimed to avoid a possible confusion.
In Chern-Simons theory at level $k'$,
many physical results, like expectation values of products of Wilson
loops, are conveniently written as functions of $k'+h$.  From the point
of view of Feynman diagram calculations, this arises because a one-loop
diagram with internal gauge bosons shifts $k'$ to  $k'+h$
\refs{\pisarski, \manny}.  
In a Hamiltonian approach to Chern-Simons gauge theory without fermions,
one sees in another way that if the parameter in the Lagrangian is
 $k'$, many physical answers
are functions of $k'+h$ \refs{\elitz,\axel}.  (This Hamiltonian approach
is much closer to what we will do in section 3
 for the theory with fermions.)
In asserting that the effective
coefficient of the Chern-Simons interaction is $k'=k-h/2$,
we are referring to an effective Lagrangian in which the fermions
have been integrated out, but one has not yet tried to solve for the
quantum dynamics of the gauge bosons.

Since the pure Chern-Simons theory is a good low energy description
for sufficiently large $k$, the 
 index of the supersymmetric theory at level $k$ can be identified
for sufficiently large $k$
with the number of supersymmetric states of the pure Chern-Simons theory
at level $k'$:
\eqn\imcob{I(k)=J(k').}

For example, suppose $G=SU(2)$.  The representations of the $SU(2)$
affine algebra at level $k'$ have highest weights of spin
$0,1/2,1,\dots,k'/2$; there are $k'+1$ such representations in all.
As $k'+1=k$ for $SU(2)$, we get
\eqn\weget{I(k)=k,}
at least for sufficiently big $k$ where the effective description
by $SU(2)$ Chern-Simons theory at level $k'$ is valid.

The formula, however, has a natural analytic continuation for all
$k$, and we may wonder if \weget\ holds for all $k$.  We will show
this in the next section by a microscopic computation, but in the meantime,
a hint that this is so is as follows.
The sign reversal $k\to -k$ is equivalent
in the Chern-Simons theory to a reversal of spacetime orientation,
so one might expect $I(-k)=I(k)$.  Actually, in general, 
the sign of the operator $(-1)^F$ in finite volume can depend on an
arbitrary choice, as in some examples in \witten.  If a parity-invariant
choice of this sign cannot be made in general, then we should
expect only $I(-k)=\pm I(k)$.  This is consistent with \weget, which
gives $I(-k)=-I(k)$.  We will see in section 3 that the general
formula, for a gauge group $G$ of rank $r$, is
\eqn\jci{I(-k)=(-1)^rI(k).}

In \weget, we can also see the claim made in the introduction:
$I(k)=0$ for $|k|<h/2$, and $I(k)\not= 0$ for $|k|\geq h/2$.
For $SU(2)$, as $h/2=1$, this is equivalent to the statement
that $I(k)$ vanishes precisely if $k=0$.
We thus learn that for $G=SU(2)$, supersymmetry is unbroken for all $k\not= 0$,
and we conjecture that it is spontaneously broken for $k=0$.
(If this is so, then in particular there is a Goldstone fermion for
$k=0$, and the pure Chern-Simons theory on which we have based
our initial derivation of \weget\ is not a good low energy description
for $k=0$.)

A similar structure holds for other groups.
For example, for $G=SU(n)$ one has
\eqn\tiun{J(k')={1\over (n-1)!}\prod_{j=1}^{n-1}(k'+j).}
(One way to compute this formula -- and its generalization to other
groups -- will be reviewed in section 3.)
When expressed in terms of $k$, this gives the formula for $I(k)$
already presented in the introduction:
\eqn\iun{I(k)={1\over (n-1)!}\prod_{j=-n/2+1}^{n/2-1}(k-j).}
We see the characteristic properties $I(-k)=(-1)^{n-1} I(k)$
and $I(k)=0$ for $|k|<n/2$.

\subsec{Microscopic Derivation Of Parity Anomaly}

The shift in the effective value of $k$ -- namely
$k'=k-\half h \,{\rm sgn}(k)$ -- has played an important role in this
discussion.  As we have already noted, the existence of this
shift implies -- since the effective Chern-Simons coupling must
be an integer -- that $k$ is congruent modulo $\Z$ to $h/2$.
When $h$ is odd -- for example, for $SU(n)$ with odd $n$ -- it
follows that $k$ is not an integer and in particular cannot be zero.
Such a phenomenon in three-dimensional gauge theories is known
as a parity anomaly \refs{\alv,\redlich}, the idea being that the theory
conserves parity if and only if $k$ vanishes, so the non-integrality
of $k$ means that parity cannot be conserved.  

The derivation of the parity anomaly from the shift in the effective
value of $k$ is valid for sufficiently large $k$ -- where there is
an effective low energy    description as a Chern-Simons theory -- but
is not valid for small $k$.  One would like to complement this
low energy explanation  by an explanation at short distances, in 
terms of the elementary degrees of freedom, that does not depend
on knowledge about the dynamics at long distances.

\def\DD{{\cal D}}

We will now review how this is done \refs{\alv,\redlich}.  In this
discussion, we assume to begin with that the gauge group $G$ is 
simply-connected (and connected), so that the gauge bundle over the
three-dimensional spacetime manifold $X$
is automatically trivial.  For most of the discussion below, the topology
of $X$ does not matter, but for eventual computation of $\Tr\,(-1)^F$,
one is most interested in $\T^3$ or $\T^2\times \R$.

The path integral in a three-dimensional gauge theory with fermions
has two factors the definition of whose phases requires care.
One is the exponential of the Chern-Simons functional.  The other
is the fermion path integral.  As the fermions are real, the fermion
path integral equals the square root of the determinant of the Dirac
operator ${\cal D}=i\Gamma\cdot D$. (When we want to make explicit
the dependence of the Dirac operator on a gauge field $A$, we write
it as $\DD_A$.  Note that we consider the massless Dirac operator.
The topological considerations of interest for the moment are independent of
the mass.) 
Thus, the factors that we must look at are
\eqn\uxxu{\sqrt{\det\,{\cal D}} \,\exp\left({ik\over 4\pi}\int \Tr\left(
A \wedge dA +{2\over 3}A\wedge A\wedge A\right)\right).}

First let us recall the issues in defining $\sqrt{\det\,{\cal D}}$.
The operator $\DD$ is hermitian, so its eigenvalues are real.
Moreover, in three dimensions, for fermions taking values in a real
bundle such as the adjoint bundle, the eigenvalues are all
of even multiplicity.  This follows from the existence of an
antiunitary symmetry analogous to CPT in four dimensions.
\foot{Use standard gamma matrix conventions such that, in a local
Lorentz frame, the gamma matrices are the $2\times 2$ Pauli spin
matrices, which are 
real and symmetric or imaginary
and antisymmetric.
The Dirac operator then commutes with the antiunitary transformation
$T:\lambda^\alpha\to \epsilon^{\alpha\beta}\bar\lambda_\beta$.
Since $T$ is antiunitary and $T^2=-1$, $\lambda$ and $T\lambda$
are always linearly independent, so the eigenstates of the Dirac
operator come in pairs.}
The determinant of the Dirac operator is defined roughly as 
\eqn\cvud{\det\,\DD=\prod_i\lambda_i,}
where the infinite product is regularized with (for example)
zeta function or Pauli-Villars regularization.  Note in particular
that the determinant is formally positive -- there are infinitely many
negative $\lambda$'s, but they come in pairs -- and this positivity
is preserved in the regularization.
Now consider the square root of the determinant, which is defined roughly
as
\eqn\hudr{\sqrt{\det \,\DD}=\prod_i{}' \lambda_i,}
where the product runs over all pairs of eigenvalues and the symbol
$\prod_i{}'$ means that (to get
the square root of the determinant) we take one
eigenvalue from each degenerate pair.  This infinite product
of course needs regularization.   Since $\det\,\DD$ has already been defined,
to make sense of $\sqrt{\det\,\DD}$ we must only define the sign.
For this we must determine, formally, whether the number of negative
eigenvalue pairs is even or odd; it is here that an anomaly will come in.

It suffices to determine the sign of $\sqrt{\det\,\DD_A}$ up to an overall
$A$-independent
sign (which cancels out when we compute correlation functions).
For this, we fix an arbitrary connection $A_0$ (chosen generically
so that the Dirac operator $\DD_{A_0}$ has no zero eigenvalues), 
and declare that
$\sqrt{\det\,\DD_{A_0}}$ is, say, positive.  Then to determine the sign
of $\sqrt{\det\,\DD_A}$ for any other connection $A$ on the same bundle, 
we interpolate from $A_0$ to $A$
via a one-parameter family of connections $A_t$, with $A_{t=0}=A_0$,
and $A_{t=1}=A$.\foot{For example,
we can take the family $A_t=tA_0+(1-t)A$.}
We follow the spectrum of $\DD_{A_t}$ as $t$ evolves
 from $0$ to 1,  and denote
 the net number of eigenvalue pairs that change sign from positive
to negative as the {\it spectral flow} $q$.  (If $A_t$ is a generic
one-parameter family, then there are no level crossings for $0\leq t\leq 1$,
and the spectral flow for $0\leq t\leq 1$ is as follows: every eigenvalue
pair flows upwards or downwards by $|q|$ units.)
Then we define the sign of 
$\sqrt{\det\,\DD_A}$ to be $(-1)^q$, the intuitive idea being that the sign
of the product in \hudr\ should change whenever an eigenvalue
pair crosses zero.  The only potential problem with this
definition is that it might depend on the path from $A_0$ to $A$.

A problem arises precisely if there is a path dependence in the value of $q$
modulo 2.  There is such path dependence if and only if
 there is a closed
path, in the space of connections modulo gauge transformations,
for which the spectral flow is odd.  To determine whether this
occurs, we proceed as follows.   Let $A_t$, for $0\leq t \leq
1$, be a family of gauge fields such that $A_1$ is gauge-equivalent
to $A_0$ by a gauge transformation $\Omega$.  
Such an $\Omega$ is classified by its ``winding number'' $\nu$
which takes values in $\pi_3(G)=\Z$.\foot{The winding number completely
specifies the topology of $\Omega$ because we are taking  
$G$ to be connected and simply-connected.  Otherwise, depending
on the topology of spacetime,
$\Omega$ may have additional topological invariants.}
In this situation,
there is a nice formula for the spectral flow.
Each $A_t$ is a connection on a trivial bundle over $X$.  
The family $A_t$, $0\leq t\leq 1$, can be fit together
to make a connection on a trivial bundle over $I\times X$, where
$I=[0,1]$ is the closed unit interval.  Gluing together the endpoints
of $I$ to make a circle $\S^1$ -- and identifying the gauge bundles
over the boundaries of $I\times X$ using the gauge transformation
$\Omega$ -- one can reinterpret the family $A_t$ as a connection on
a possibly nontrivial bundle $E$ over $\S^1\times X$.  This bundle
has  instanton number $\nu$, determined by the topological twist
of $\Omega$.  The spectral flow is then
\eqn\oppo{q= h \nu.}
This relation between spectral flow and the topology of the bundle
\ref\aps{M. F. Atiyah, V. K. Patodi, and I. M. Singer,
``Spectral Asymmetry And Riemannian Geometry, I, II, III,''
Proc. Camb. Philos. Soc. {\bf 77} (1975) 43, {\bf 78} (1980)
2848, {\bf 79} (1976) 1.}, which is important in instanton
physics \ref\cgdjr{G. 't Hooft, ``Symmetry Breaking Through
Bell-Jackiw Anomalies,'' Phys. Rev. Lett. {\bf 37} (1976) 8;
D. J. Gross, C. G. Callan, Jr., and R. F.
 Dashen, ``The Structure Of The Gauge Theory Vacuum,''
 Phys. Lett. {\bf 63B} (1976) 334; R. Jackiw and C.  Rebbi,
 ``Vacuum Periodicity In A Yang-Mills Quantum Theory,'' Phys. Rev.
 Lett. {\bf 37} (1976) 172.},
 is proved roughly as follows
using the index theorem for the four-dimensional
Dirac operator on $\S^1\times X$.  We call that operator $\DD_4$ and
let $\DD_{4,+}$ and $\DD_{4,-}$
be the restrictions of $\DD_4$ to spinors of positive
or negative chirality, namely
\eqn\moppo{\DD_{4,\pm} =\pm {\partial\over\partial t}+ \epsilon \DD_{A_t},}
where  $\epsilon$ is a positive real number that one can
introduce by scaling the metric on $X$.  For small  $\epsilon$,
the Dirac equation $\DD_{4,\pm}\psi=0$ can be studied in terms of the
$t$-dependent spectrum of $\DD_{A_t}$.  If $\lambda_i(t)$ are the
eigenstates of $\DD_{A_t}$ with eigenvalues $s_i(t)$, then 
one can approximately solve the four-dimensional Dirac equation with
the formula
\eqn\oxno{ \Psi_i(t) =\sum_{k\in \Z}\exp\left(
\mp \epsilon\int_0^{t+k}s_i(t')dt'\right)
\lambda_i(t+k)}
where the sign in the exponent is $\mp$ to give zero modes of $\DD_{4,\pm}$.
The sum over $k$ has been included
to  ensure $\Psi_i(t+1)=\Psi_i(t)$.
Different $\lambda_i$ that are related by spectral flow (that is by
$t\to t+1$) give the same $\Psi_i$, so for generic spectral flow 
 there are $|2q|$ linearly independent
four-dimensional solutions of this kind.  For $\Psi_i$ to be square 
integrable, the exponential factor in \oxno\ must 
vanish for  $t'\to\pm \infty$,
so $\mp s_i(t')$ must be negative for $t'\to \infty$ and also
for $t'\to -\infty$.
This determines that the chirality of the solutions is the same as the
sign of the spectral flow $q$.
The upshot is that the index $I(\DD_4)$ of $\DD_4$ equals $2q$.  (The factor
of 2 arises 
because we defined $q$ by counting pairs of eigenvalues; each pair
contributes two four-dimensional zero modes.)
  On the other hand, the index theorem for the Dirac operator
gives 
$I(\DD_4)= 2 h \nu$.  Combining the formulas for $I(\DD_4)$ gives \oppo.

Now we put our results together.
Under the gauge transformation $\Omega$, or in other words in interpolating
from $t=0$ to $t=1$, the sign of $\sqrt{\det\,\DD}$ changes
by $(-1)^q$.  In view of \oppo, this
factor is $(-1)^{h\nu}$.  On the other hand, the change in the
Chern-Simons functional under a gauge transformation of winding number
$\nu$ is $2\pi \nu$.  So under the gauge transformation $\Omega$,
the dangerous factors \uxxu\ in the path integral pick up a factor
\eqn\ixnc{(-1)^{h\nu}\exp(2\pi i k\nu).}
Gauge invariance of the theory amounts to the statement that
this factor must be an integer for arbitrary integer $\nu$,
and this gives us the restriction on $k$:  
\eqn\pixnc{k\cong {h\over 2} ~{\rm modulo}~\Z.}

\subsec{The $SU(n)/\Z_n$ Theory}

Now we have assembled the ingredients to put the hypothesis
of dynamical supersymmetry breaking for $|k|<h/2$ to an apparently
rather severe test.  The discussion is most interesting for the case
$G=SU(n)$, so we focus on that case.

The idea is to consider $\Tr\,(-1)^F$ for an $SU(n)/\Z_n$ theory
on $\T^2$.  The key difference between $SU(n)$ and $SU(n)/\Z_n$
is that any $SU(n)$ bundle on $\T^2$ is trivial, but an $SU(n)/\Z_n$
bundle on $\T^2$ is characterized by a ``discrete magnetic flux'' $w$ 
that takes values in $\Z_n$.  (For $n=2$, $SU(2)/\Z_2=SO(3)$,
and the discrete flux is the second Stieffel-Whitney class of the bundle.)
An example of a bundle with any required value of $w$ is as follows.
Consider a flat $SU(n)/\Z_n$ bundle whose holonomies $U$ and $V$ around the
two directions in $\T^2$, if lifted to $SU(n)$, obey
\eqn\kxik{UV=VU\exp(2\pi i r/n).}
Such a flat bundle has $w=r$.

The computation of $\Tr\,(-1)^F$ for this theory can be made
very easily in case $r$ and $n$ are relatively prime, for instance $r=1$.
(The computation can be done for any $r$ by using the relation
to the WZW model of $SU(n)/\Z_n$, along the lines of section 2.1
above, or more explicitly using the techniques of section 3.)
The idea is simply \witten\ that zero energy quantum states
are obtained, for weak coupling, by quantizing the space of zero
energy classical states (including possible bosonic or fermionic
zero modes).  A zero energy classical configuration of the gauge
fields is a flat connection.  For $r$ and $n$ relatively prime,
a flat connection -- that is a pair of matrices $U$ and $V$ obeying
\kxik\ -- is unique up to gauge transformation.  Moreover, in expanding
around such a flat connection, there are no bosonic or fermionic zero modes.
Hence, the quantization is straightforward: quantizing a unique, isolated
classical state of zero energy, with no zero modes, gives a unique
quantum state.\foot{There is actually a potential subtlety in this statement,
though it is  inessential in the examples under discussion.  One must
verify that the one state in question obeys Gauss's law, in other words
that it is invariant under the gauge symmetries left unbroken
by the classical solution that is being quantized.}
  The index is therefore $\pm 1$ (with the sign possibly
depending on a choice of sign in the definition of the operator
$(-1)^F$).

Note that $k$ plays no role in this argument.  Hence, for any $k$
for which the $SU(n)/\Z_n$ theory exists, this theory, if formulated
on a bundle with $r$ prime to $n$, has a supersymmetric vacuum state
for any volume of $\T^2$.  Taking the limit of infinite volume,
it follows that the $SU(n)/\Z_n$ theory, for any such $k$, has zero
vacuum energy and hence unbroken supersymmetry.

But in infinite volume, the $SU(n)$ and $SU(n)/\Z_n$ theories
are equivalent.\foot{Except for questions of which operators one
chooses to probe them by; such questions are irrelevant for the present
purposes.}  Hence for any  $k$ for which the $SU(n)/\Z_n$ theory is
defined, the $SU(n)$ theory has unbroken
supersymmetry.

Does this not disprove the hypothesis that the $SU(n)$ theory
has spontaneously broken supersymmetry in the ``gap,'' that is
for $|k|<n/2$?  In fact, there is an elegant escape which we will
now describe.

The allowed values of $k$ were determined for $SU(n)$ by requiring that
\eqn\xon{(-1)^{n\nu}\exp(2\pi i k \nu)}
should equal 1 for all integer values of the instanton number $\nu$.
(We have rewritten \ixnc\ using the fact that $h=n$ for $SU(n)$.)
For $SU(n)/\Z_n$, there is a crucial difference: the instanton number $\nu$
is not necessarily an integer, but takes values in $\Z/n$
\ref\otherthooft{G. 't Hooft, ``Some Twisted Self-Dual Solutions For
The Yang-Mills Equations On A Hypertorus,'' Commun. Math. Phys. {\bf B81}
(1981) 267.}.
(For example, setting $X=\T^3$,
an $SU(n)/\Z_n$ bundle on $\S^1\times \T^3=\T^4$
that has unit magnetic flux in the 1-2 and 3-4 directions and
other components vanishing has instanton number $1/n$ modulo $\Z$.
In fact, on a four-manifold that is not spin, the instanton number
takes values in $\Z/2n$, but for our present purpose -- as the supersymmetric
theory has fermions -- only spin manifolds are relevant.)
Hence gauge invariance of the theory requires that \xon\ should equal
1 not just for all $\nu\in \Z$, but for all $\nu\in \Z/n$.
This gives the relation 
\eqn\ujulp{k\cong {n\over 2} ~{\rm modulo}~n.}
Thus, for $SU(n)/\Z_n$, $k$ cannot be in the ``gap'' $|k|<n/2$,
and the behavior of the $SU(n)/\Z_n$ theory in finite volume
 cannot be used to exclude the hypothesis
that in the gap supersymmetry is dynamically broken. 
Though this does not prove that supersymmetry is broken in the gap for $SU(n)$,
the elegant escape does suggest that that is the right interpretation.

\newsec{Microscopic Computation Of The Index}

In this section, we will  make a microscopic computation of
$\Tr\,(-1)^F$ in the $N=1$ supersymmetric pure gauge theory
in three spacetime dimensions.  We consider first the case that the
gauge group $G$ is simply connected.

\def\M{{\cal M}}
We thus formulate the theory on a spatial torus $\T^2$ (times time) and
look for zero energy states.  As
in \witten, the computation will be done by a ``Born-Oppenheimer 
approximation,'' quantizing the space of classical zero energy
states, and is valid for weak coupling or small volume of $\T^2$.
To be more exact, we work on a torus of radius $r$, and let $e$ and $k$
denote, as before, the gauge and Chern-Simons couplings.  Particles
with momentum on $\T^2$ have energies of order $1/r$, while the
fermion and gauge boson bare mass is $e^2k$.  We work in the region
\eqn\jiko{e^2k<<{1\over r}.}
We will write an effective Hamiltonian that describes states with
energies of order $e^2k$ (or less) but omits states with energies of order
$1/r$.

A zero energy classical gauge field configuration 
is a flat connection and is determined up to gauge transformation
by its holonomies $U,V$ around the two directions in $\T^2$.  These
holonomies, since they commute, can simultaneously be conjugated
to the maximal torus $\U$ of $G$,\foot{The analogous statement
can fail -- see the
appendix of \ref\uggywitten{E. Witten, ``Toroidal Compactification Without
Vector Structure,'' JHEP {\bf 9802} (1998) 006, hep-th/9712028.} 
-- for the case of three commuting elements
of $G$.}
 in a way that is unique up to
a Weyl transformation.  The moduli space $\M$ of flat 
$G$-connections on $\T^2$ is thus a copy of $(\U\times \U)/W$, where
$W$ is the Weyl group.  
Concretely, a flat connection on $\T^2$ can be represented by
a constant gauge field
\eqn\imoco{A_i=\sum_{a=1}^rc_i^aT^a,}
where the $T^a,\,a=1,\dots,r$ are 
a basis of the Lie algebra of $\U$ and the $c_i^a$ can be regarded
as constant abelian gauge fields on $\T^2$.  The flat metric on $\T^2$
determines a complex structure on $\T^2$; it also determines a complex
structure on $\M$ in which the complex coordinates are the components
$c_{\bar z}^a$ of the one-forms $c_i^a$.  The $c_i^a$ are defined
modulo $2\pi$ shifts.

To construct the right quantum mechanics on $\M$, we must also 
look at the fermion zero modes.  
Actually, by ``zero modes'' we mean modes whose energy is at most
of order $e^2k$, rather than $1/r$.  In finding
these modes, we can ignore the fermion bare mass and look for zero
modes of the massless two-dimensional Dirac operator $\DD$.  We then
will write an effective Lagrangian that incorporates the effects of
the fermion bare mass.
Let $\lambda_+$ and $\lambda_-$
be the gluino fields of positive and negative chirality on $\T^2$.
(They are hermitian conjugates of each other.)  For a diagonal
flat connection such as we have assumed, the 
 equation $\DD \lambda=0$ has a very simple structure.
For generic $U$ and $V$ (or equivalently for generic $c_i^a$),
the ``off-diagonal'' fermions have no zero modes, while the ``diagonal''
fermions have ``constant'' zero modes.  In other words, the zero
modes of $\lambda_\pm$ are given by the ansatz
\eqn\xonxon{\lambda_\pm =\sum_{a=1}^r\eta_\pm^a T^a,}
with $\eta_\pm^a$ being anticommuting constants.

Now let us discuss quantization of the fermions.  Quantization of the
nonzero modes gives a Fock space.  Quantization of the
zero modes $\eta_\pm^a$ is, as usual, more subtle.
The canonical anticommutation relations of the $\eta$'s are,
with an appropriate normalization,
\eqn\ubuxxon{\{\eta_+^a,\eta_-^b\}=\delta^{ab},~~\{\eta_+,\eta_+\}=\{\eta_-,
\eta_-\}=0.}
Thus, we can regard the $\eta_+$ and $\eta_-$ as creation and annihilation
operators.  For example, we can introduce a state $|\Omega_-\rangle$
annihilated
by the $\eta_-^a$, and build other states by acting with $\eta_+^a$'s;
or we can introduce a state $|\Omega_+\rangle$ annihilated
by the $\eta_+^a$'s, and build the rest of the Hilbert space by
acting with $\eta_-^a$'s.
The relation between the two descriptions is of course
\eqn\sxni{|\Omega_+\rangle =\prod_{a=1}^r\eta_+^a|\Omega_-\rangle.}

Now let us try to define the operator $(-1)^F$.  It is clear how
we want $(-1)^F$ to act on the Fock space built by quantizing the
nonzero modes of the fermions: it leaves the ground state invariant
and anticommutes with all nonzero modes of $\lambda$.  The only
subtlety is in the action on the zero mode Hilbert space.  There
is in general no natural choice for the sign of $(-1)^F$.  If
we pick, say $(-1)^F|\Omega_-\rangle=+|\Omega_-\rangle$, then
\sxni\ implies that $(-1)^F|\Omega_+\rangle=(-1)^r|\Omega_+\rangle$.
Thus, if $r$ is even, we can fairly naturally pick both of the
states $|\Omega_\pm\rangle$
to be bosonic.  But if $r$ is odd, then inevitably one is bosonic
and one is fermionic; which is which depends on a completely arbitrary
choice.  Now we can explain an assertion in section 2, namely that
\eqn\ucxu{I(-k)=(-1)^rI(k).}
Changing the sign of the Chern-Simons level is equivalent to a transformation
that reverses the orientation of $\T^2$.  Such a transformation exchanges
$\lambda_+$ with $\lambda_-$, and so exchanges $|\Omega_+\rangle$
with $|\Omega_-\rangle$.  This exchange reverses the sign of the $(-1)^F$
operator if $r$ is odd, and that leads to \ucxu.

The Hilbert space made by quantizing the fermion zero modes has a very
natural interpretation.  
The quadratic form \ubuxxon\ has the same structure as the metric
$ds^2= \sum_{a,b}\delta_{ab} dc_z^a\otimes dc_{\bar z}^b$ 
of $\cal M$, so the $\eta$'s
can be interpreted as gamma matrices on $\M$.  Hence the Hilbert
space obtained by quantizing the zero modes is the space of spinor
fields on $\M$, with values in a line bundle ${\cal W}$
that we have not yet identified.  (Such a line bundle may appear because,
for example,
for a given point on ${\cal M}$, the state $|\Omega_+\rangle$ is unique
up to a complex multiple, but as one moves on ${\cal M}$, it varies
as the fiber of a not-yet-determined complex line bundle.)

Because $\M$ is a complex manifold,  spinor fields on $\M$
have a particularly simple description.  Let $K$ be the canonical
line bundle of $\M$, and assume for the time being that there exists
on $\M$ a line bundle $K^{1/2}$.  Then the space of spinors on $\M$ is the
same as the space $\Omega^{0,q}(M)\otimes K^{1/2}$ of $(0,q)$-forms
on $\M$ (for $0\leq q \leq s$) with values in $K^{1/2}$.
In this identification, we regard $|\Omega_+\rangle$ as a $(0,0)$-form
on $\M$ (with values in a line bundle), and we identify
a general state in the fermionic space
\eqn\ixnin{\eta_-^{a_1}\dots \eta_-^{a_q}|\Omega_+\rangle}
as a $(0,q)$ form on $\M$.  From this point of view, we identify
$\eta_-^{a}$ with the $(0,1)$ form $dc_{\bar z}^a$, and $\eta_+^a$
with the ``contraction'' operator that removes the one-form $dc_{\bar z}^a$
from a differential form, if it is present.  (Of course, by exchanging
the role of $\eta_-$ and $\eta_+$, we could instead regard the spinors on $\M$
as $(q,0)$-forms, with values in a line bundle.)  

\def\L{{\cal L}}
\def\W{{\cal W}}
The relation of spinors on $\M$ to $(0,q$)-forms has the following
consequence.
The Dirac operator $\DD$ acting on sections of a holomorphic
line bundle $\W$ over a complex manifold $\M$
has a decomposition
\eqn\ixxix{\DD=\bar\partial+\bar\partial^\dagger,}
where $\bar\partial$ is the $\bar\partial$ operator acting on
$(0,q)$-forms with values in $\W\otimes K^{1/2}$, and $\bar\partial^\dagger$
is its adjoint.
These operators obey
\eqn\mocobi{\{\bar\partial,\bar\partial^\dagger\}=H,~~\bar\partial^2
=(\bar\partial^\dagger)^2=0,}
where $H=\DD^2$.
If we identify $\bar\partial$ and $\bar\partial^\dagger$ with
the two supercharges and $H$ with the Hamiltonian, then \mocobi\ coincides
with the supersymmetry algebra of a $2+1$-dimensional system with
$N=1$ supersymmetry, in a sector in which the momentum vanishes.
In the present discussion the momentum vanishes because the classical zero energy
states that we are quantizing all have zero momentum.  This strongly
suggests that, in the approximation of quantizing the space of classical
zero energy states, the supercharges reduce to (a multiple of)
$\bar\partial$ and
$\bar\partial^\dagger$.

It is not difficult to show this and at the same time identify
the line bundle ${\cal W}$.  In canonical quantization of the Yang-Mills
theory with Chern-Simons coupling, the momentum conjugate to $A_i^a$
is
\eqn\ixono{\Pi_i^a={F_{0i}^a\over e^2}-{k\over 4\pi}\epsilon_{ij}A_j^a.}
Writing formally $\Pi_i^a=-i\delta/\delta A_i^a$, we
have
\eqn\kixono{{F_{0i}^a(x)\over e^2} =-i{D\over DA_i^a(x)},}
where $D/DA_i^a(x)$ is a ``covariant derivative in field space,''
\eqn\lopo{{D\over DA_i^a(x)}={\delta\over \delta A_i^a(x)}+i{k\over 4\pi}
\epsilon_{ij}A_j^a.}
The object $D/DA_i^a$ is a connection on a line bundle $\W$ over the
space of connections.  The connection form of $\W$ is
$(k/4\pi)\epsilon_{ij}A^j$.  Requiring that the curvature
form of $\W$ should have periods that are integer multiples of $2\pi$
gives a condition that is equivalent to the quantization \templeton\
of the Chern-Simons coupling (see \refs{\elitz - \cresci} for more on such
matters), so if we set $k=1$ the line bundle
that we get is the most basic line bundle ${\cal L}$ over the phase space,
in the sense that it has positive curvature and all other line bundles
over the phase space are of the form ${\cal L}^n$ for some integer $n$.
The factor of $k$ in \lopo\ means that the line bundle $\W$ 
over the phase space $\M$  is $\W=\L^k$.  So the states are spinors
with values in $\L^k$, or equivalently $(0,q)$-forms with values
in $\L^k\otimes K^{1/2}$.

As for the supercharges, they
are 
\eqn\truto{Q_\alpha={1\over e^2}\int_{\T^2}\,\Gamma^{IJ}_{\alpha\beta}\,
\Tr F_{IJ}\lambda^\beta.}
To write an effective formula in the space of zero energy states,
we set the spatial part of $F$ to zero. The supercharges $Q_\pm$ 
of definite two-dimensional chirality then become
\eqn\plxon{\eqalign{Q_-& ={1\over e^2}\int_{\T^2}\Tr\,F_{0 \bar z} \lambda_+
                              = \int_{\T^2}\Tr\lambda_+{D\over DA_{ z}}  \cr
                    Q_+& ={1\over e^2}\int_{\T^2}\Tr\,F_{0 z}\lambda_-
                                =\int_{\T^2}\Tr\lambda_-{D\over DA_{\bar z}}.
                                \cr}}        
Evaluating this expression in the space of zero modes, 
the $\lambda$'s become gamma matrices (or raising and lowering operators)
on spinors over the moduli space $\M$; and $D/DA_{\bar z}$ and $D/DA_z$
are holomorphic and antiholomorphic covariant derivatives on $\M$.
Altogether, the supercharges $Q_-$ and $Q_+$ 
 reduce to $e$ times the $\bar\partial$ and $\bar\partial^\dagger$ operators
on spinors valued in $\W=\L^k$.\foot{The factor of $e$ arises because
the $\lambda$ kinetic energy in the original Lagrangian was
$\bar\lambda \DD\lambda/e^2$, so $\lambda/e$ is a canonically
normalized fermion.  As the supercharges are properly normalized
as $e\bar\partial$ and $e\bar\partial^\dagger$, the Hamiltonian
is $H=e^2\{\bar\partial,\bar\partial^\dagger\}$.}

In this discussion,  we have not
incorporated explicitly the fermion bare mass $e^2k$.  But that bare
mass is related by supersymmetry to the Chern-Simons coupling, which
we have incorporated, so the supersymmetric effective Hamiltonian
$H=\{\bar\partial,\bar\partial^\dagger\}$ that we have written inevitably
includes the effects of the fermion bare mass.  This arises as follows:
because there is a ``magnetic field'' on $\M$ proportional to $k$
(with connection
form on the right hand side of \lopo), the operator $H=e^2\{\bar\partial,
\bar\partial^\dagger\}$, if written out more explicitly, contains
a term $e^2k\eta_-^a\eta_+^b\delta_{ab}$.
This coupling is the bare mass term, written 
in the space of $\eta$'s.

\subsec{Calculations}

Now we will perform calculations of $\Tr\,(-1)^F$.  First
we consider the case that $G=SU(n)$.  

For $SU(n)$, the moduli space $\M$ is a copy of ${\bf CP}^{n-1}$.
\foot{For example, for $n=2$, the maximal torus $\U$ is a circle
and the Weyl group is $W=\Z_2$, so $\M=(\U\times \U)/W=\T^2/\Z_2$,
which is an orbifold version of $\S^2={\bf CP}^1$.  For general $n$,
the standard proof that $\M={\bf CP}^{n-1}$ can be found,
for example, in section 2.1 of \ref\elliptic{R. Friedman, J. Morgan,
and E. Witten, ``Vector Bundles And $F$ Theory,'' Commun. Math. Phys.
{\bf 187} (1997) 679, hep-th/9701166.}.}
The basic line bundle over $\M$ is $\L={\cal O}(1)$, the bundle
whose sections are functions of degree one in the homogeneous coordinates
of ${\bf CP}^{n-1}$.  The canonical bundle of ${\bf CP}^{n-1}$ is
$K=\L^{-n}$.

The quantum Hilbert space found in the above Born-Oppenheimber
approximation  is the space of spinors with values in $\W=\L^k$,
or equivalently $(0,q)$-forms with values in $\W\otimes K^{1/2}
=\L^{k-n/2}$.  Since only integral powers of $\L$ are well-defined
as line bundles over ${\cal M}$, we get the restriction
\eqn\mxinn{k\cong {n\over 2}~{\rm modulo}~\Z.}
This is the restriction found in \who\ and reviewed in section 2; we have
now given a Hamiltonian explanation of it.

Since the supersymmetry generators are the $\bar\partial$ and $\bar\partial
^\dagger$ operators, the space of supersymmetric states, in this
approximation, is
\eqn\xnonncpo{\oplus_{i=0}^n H^i({\bf CP}^{n-1},{\cal L}^{k-n/2}).}
The supersymmetric index is
\eqn\xocni{I(k)=\sum_{i=0}^n(-1)^i{\rm dim}\,H^i({\bf CP}^{n-1},\L^{k-n/2}).}
This can be computed by a Riemann-Roch formula, which implies in
particular that $I(k)$ is a polynomial in $k$ of order $n$.

However, for a more precise description -- and in particular to see
supersymmetry breaking in the ``gap,'' $|k|<n/2$ -- we wish to compute
the individual cohomology groups, and not just the index.
For this computation, see for example
\ref\compref{R. Hartshorne, {\it Algebraic Geometry} (Springer-Verlag,
1977), section III.5.}.
For $-n<t<0$, one has $H^i({\bf CP}^{n-1},\L^t)=0$ for all $i$.
Hence, for
\eqn\xonnc{-{n\over 2} < k < {n\over 2},}
there are no  zero energy states at all in the present approximation.
Thus, for this range of $k$, supersymmetry is spontaneously broken
if the theory is formulated on a two-torus with sufficiently weak
coupling that our analysis is a good approximation.  (Because the ground
state energy in finite volume
is a real analytic function of the volume, it also follows
that supersymmetry is unbroken for any generic volume on $\T^2$.)
 This hints but
certainly does not prove that also for infinite volume, supersymmetry
is spontaneously broken if $|k|<n/2$.

For $t\geq 0$, $H^i({\bf CP}^{n-1},\L^t)=0$ for $i>0$, and
$H^0({\bf CP}^{n-1},\L^t)$ is the space of homogeneous polynomials
of degree $t$ in the $n$ homogeneous coordinates of ${\bf CP}^{n-1}$.
The dimension of this space is $n(n+1)\dots (n+t-1)/t!=(n+t-1)!/t!(n-1)!$  
Setting
$t=k-n/2$ and interpreting this dimension as the supersymmetric
index $I(k)$, we  get the formula for
$I(k)$ that was stated in the introduction:
\eqn\kxnno{I(k) = {1\over (n-1)!}\prod_{j=-n/2+1}^{n/2-1}(k-j).}

Finally, 
Serre duality determines what happens for $t\leq -n$ in terms of the
results for $t\geq 0$.  In particular,
for $t\leq -n$, the cohomology $H^i({\bf CP}^{n-1},\L^t)$
vanishes except for $i=n-1$, and is dual to $H^0({\bf CP}^{n-1},\L^{-n-t})$.
From this, we get a formula for $I(k)$ with $k\leq -n/2$ which
coincides with \kxnno.  Note that for $k\geq n/2$,
all supersymmetric states are bosonic, and for $k\leq -n/2$, all
supersymmetric states have statistics $(-1)^{n-1}$.
Serre duality gives directly $I(-k)=(-1)^{n-1}I(k)$.

\bigskip\noindent
{\it Generalization To Other Groups}

We will now more briefly summarize the generalization for an arbitrary simple,
connected and simply-connected gauge group $G$ of rank $r$.

First of all, the moduli space $\M$ is a weighted projective
space ${\bf WCP}^r_{s_0,s_1,\dots,s_r}$, where the weights $s_i$ are 1 and
the coefficients of the highest coroot of $G$.  This is a theorem of Looijenga;
for an alternative proof see \elliptic. 
In particular, the weights obey
$\sum_{i=0}^rs_i=h$.  The basic line bundle over $\M$ is ${\cal L}=
{\cal O}(1)$, characterized by the fact that sections of ${\cal L}^t$
for any $t$ are functions of weighted degree $t$ in the homogeneous coordinates
of $\M$.  The canonical bundle of $\M$ is $K=\L^{-h}$.
So, in the Born-Oppenheimer approximation, the low-lying states
are spinors valued in $\L^{k-h/2}$.  Integrality of the exponent
gives again the result that $k-h/2$ must be integral.

The space of supersymmetric states is, again,
\eqn\xonnco{\oplus_{i=0}^rH^i(\M,\L^{k-h/2}).}  
A weighted projective space has certain properties in common with
an ordinary projective space.  One of these is that $H^i(\M,\L^t)=0$
for all $i$ if $-h<t<0$.  This implies (in finite volume)
supersymmetry breaking in the ``gap,'' $|k|<h/2$. 
For $t\geq 0$, the cohomology groups vanish except in dimension
0, and $H^0(\M,\L^{t})$ is the space of polynomials homogeneous and
of weighted degree $t$ in the homogeneous coordinates of $\M$.
In particular, $I(k)>0$ for $k\geq h/2$, and supersymmetry is unbroken.
Serre duality asserts that $H^i(\M,\L^t)$ is dual to
$H^{r-i}(\M,\L^{-h-t})$, and
relates the region $k\leq -h/2$ to $k\geq h/2$.
In particular, 
for $k\leq h/2$, the only nonzero cohomology group is in dimension $r$,
the states have statistics $(-1)^r$, and the index is determined by
$I(-k)=(-1)^rI(k)$ and so is in particular nonzero.

\subsec{Orbifolds And Anyons}

Here, we make a few miscellaneous comments on
the problem.

The moduli space $\M$ is an orbifold $\M=(\U\times \U)/W$,
a quotient of a flat manifold by a finite group.  However, we have
not used this fact in computing the index. The reason is that although
the moduli space $\M$ is an orbifold, the quantum mechanics on $\M$ is
not orbifold quantum mechanics, that is, it is not obtained from
supersymmetric free particle motion on $\U\times \U$ by 
imposing $W$-invariance.
Rather, the quantum mechanics on $\M$ depends on the line bundle $\L^k$.

One can ask whether there are values of $k$ at which the quantum
mechanics on $\M$ reduces to orbifold quantum mechanics.  
We will approach this as follows.
 We begin with a system consisting of $(0,q)$-forms on
$\U\times \U$ with the Hamiltonian being simply the Laplacian
(relative to the flat metric on $\U\times \U$).  In orbifold
quantum mechanics, we want $W$-invariant states of zero energy.

A zero energy state
must have a wave function that is invariant under translations on
$\U\times \U$.  This means that the bosonic part of the wave-function,
being a constant function, is $W$-invariant.  Hence, $W$-invariance
must be imposed on the fermionic part of the wave function, which
as we recall takes values in
 a fermionic  Fock space with basis obtained by acting
with creation operators on a vacuum $|\Omega_-\rangle $ or $|\Omega_+\rangle$.

The states $|\Omega_-\rangle$ and $|\Omega_+\rangle$ transform as
one-dimensional representations of $W$, since the condition
that a state be annihilated by all $\eta_+^a$'s (or by all
$\eta_-^a$'s) is  Weyl-invariant.  The group $W$ has two one-dimensional
representations: the trivial representation; and a representation $R$ in
which each elementary reflection acts by $-1$.  
Since
\eqn\guffu{|\Omega_+\rangle =\prod_{a=1}^r\eta_+^a|\Omega_-\rangle,}
and the product $\prod_{a=1}^r\eta_+^a$ is odd under every elementary
reflection, the two states $|\Omega_-\rangle$ and $|\Omega_+\rangle$
transform oppositely: one transforms in the trivial representation of $W$,
and the other transforms as $R$.  

Suppose that we take the $W$ action such that $|\Omega_-\rangle$
transforms trivially.  
Then $|\Omega_-\rangle$ itself (times a constant function on $\U\times\U$)
is a $W$-invariant state of zero energy, and is in fact the only one.
To prove the uniqueness, one can use the fact that
 the $W$ action on the fermion Fock space
is the same as that on the $(0,q)$-forms on $\U\times \U$.  The $W$-invariant
and translation-invariant states on $\U\times \U$ can therefore
be identified with the $\bar\partial$ cohomology
group $H^i((\U\times\U)/W,{\cal O})$, where 
${\cal O}$ is a trivial holomorphic line bundle.  
This cohomology is one-dimensional for $i=0$ and vanishes for $i>0$,
since $(\U\times \U)/W={\bf CP}^{n-1}$.

Now let us compare this orbifold quantum mechanics to the Born-Oppenheimer
quantization of the gauge theory.
In the latter, at general level $k$,
we identified the supersymmetric states with elements of
$H^i(\M,\L^{k-h/2})$, where
$\M=(\U\times\U)/W$.  This agrees with the orbifold answer
$H^i(\M,{\cal O})$ if and only if $k=h/2$, so that
must be the correct value of $k$ corresponding to orbifold quantum
mechanics with $|\Omega_-\rangle$ assumed to be Weyl-invariant.
The other possibility, that $|\Omega_+\rangle$ is Weyl-invariant,
is obtained by reversal of orientation, which is equivalent to $k\to -k$;
so this other orbifold quantum mechanics should correspond to $k=-h/2$.

These arguments strongly suggest that the low energy quantum mechanics
is just orbifold quantum mechanics for these special values of $k$.
As we discuss below and
in section 4, these are apparently the values for which
the theory is confining.

\bigskip\noindent{\it Anyons}

It is perhaps surprising that the ``simple'' orbifold cases correspond
not to the obvious case
$k=0$ but to $k=\pm h/2$.  Let us see instead consider what happens for
$k=0$.  For simplicity, we take $G=SU(2)$, so that $\U$ is a circle
and $W=\Z_2$. 

Even though $\M=(\U\times \U)/\Z_2$ is an orbifold, the quantum
mechanics is, as we have seen, not orbifold quantum mechanics for $k=0$.
To measure the failure, let us see what happens near the orbifold
singularities of $(\U\times \U)/\Z_2$.  For example, we can consider
the singularity associated with the trivial flat connection, where the
$c_i$ introduced in \imoco\ all vanish.  (For $SU(2)$, the index $a$ takes
only one value, so we write the $c_i^a$ just as $c_i$.)  
The Weyl group acts as $c_i\to -c_i$, and there is a singularity
at $c_i=0$.  How can we best understand this singularity?  Near
$c_i=0$, it is more illuminating to consider not compactification
from $2+1$ dimensions to $0+1$ -- as we have done so far -- but dimensional
reduction to $0+1$ dimensions, in which one starts with $2+1$-dimensional
super Yang-Mills thoery and by fiat one requires the fields
to be invariant under spatial translations.  Dimensional reduction
and compactification differ in that the compactified theory also
has modes of non-zero momentum along $\T^2$, and has periodic identifications
of the $c_i$ under $c_i\to c_i+2\pi$.  These are irrelevant for
studying the singularity near $c_i=0$.

In the dimensionally reduced theory, there is a $U(1)$ symmetry under rotations
of the $c_1-c_2$ plane.  (The compactified theory only has a discrete
subgroup of this symmetry; that is one of the main reasons to consider
the dimensionally reduced theory in the present discussion.)  We will
call the generator of this $U(1)$ the angular momentum.
The fermion Fock space has, for $SU(2)$, only the
 two states $|\Omega_-\rangle$ and
$|\Omega_+\rangle$ (as there is only one creation operator and one
annihilation operator).  Since a fermion creation operator, of spin $1/2$,
maps one to the other, their angular momenta 
are $j$ and $j-1/2$ for some $j$.  But 
the dimensionally reduced theory has also a parity symmetry
exchanging these two states and reversing the sign of the angular
momentum.  Hence $j=1/4$, and the two states have angular momenta
$1/4,-1/4$.  This contrasts with  orbifold quantum mechanics on $\R^2/\Z_2$,
where the spins are half-integral. Thus a precise measure of
the difference of the $k=0$ system from an orbifold is that the
$k=0$ system generates ``anyons,'' states whose angular momentum
does not take values in $\Z/2$.

Such anyons may arise in the compactification of Type IIB
superstring theory to three dimensions on a seven-manifold $X$ of
$G_2$ holonomy.  Consider a system of $m$ parallel sevenbranes
wrapped on $X$.  This system is governed by $2+1$-dimensional
$U(m)$ super Yang-Mills
theory with two supercharges, dimensionally reduced to $0+1$ dimensions.
This suggests that the wrapped sevenbranes may be anyons of spins
$\pm 1/4$, but to be certain, one would need to look closely at the
definition of angular momentum for these string theory excitations.

\subsec{Discrete Electric And Magnetic Flux}

We will briefly discuss the generalization of the computation
to incorporate discrete electric and magnetic flux.  (We will
consider only the simplest versions of electric or magnetic flux.
It is of course possible to mix the two constructions.)

Including magnetic flux simply means taking the gauge group $G$
not to be simply connected and working on a non-trivial gauge bundle
$E$ over $\T^2$.  The moduli space $\M$ of zero energy gauge configurations
is now the moduli space of flat connections on $E$.  In certain cases,
mentioned in section 2.2 above, $\M$ is a single point, and the
quantization is then completely straightforward.  In general,
$\M$ is always a weighted projective space (which can be constructed
using the technique in \elliptic), and the quantum ground states
are always, as above, $H^i(\M,\L^{k-h/2})$.  In particular,
the index is always nonvanishing for $|k|\geq h/2$.

Including electric flux means that one goes back to the case that
$G$ is simply connected.  One considers a gauge 
transformation $U$ that, in going around, say, the first circle in 
$\T^2=\S^1\times \S^1$, 
transforms as 
\eqn\ixon{U\to U\omega,} 
where $\omega$ is an element of the
center of $G$.   
This transformation is a symmetry $T_\omega$ of the theory.  If, for example,
$\omega$ is of order $s$, then $U^s$ generates a gauge transformation
that is homotopic to the identity, and $T_\omega^s=1$ on all physical
states.  The eigenvalues of $T_\omega$ are thus of the form
$\exp(2\pi ir/s)$
where $r$ is an integer called the discrete electric flux.

A simple way to determine the action of $T_\omega$ on the space
${\cal H}$ of supersymmetric ground states of our supersymmetric gauge theory
is to use the relation
of pure Chern-Simons theory at level $k'$ 
to the WZW model, also at that level.  The Hilbert space of
the pure Chern-Simons theory in quantization on $\T^2$ has a basis
that can be described as follows.  Regard $\T^2=\S^1\times \S^1$
as the boundary of $\S^1\times D$, where $D$ is a two-dimensional
disc.  Consider the Chern-Simons path integral on $\S^1\times D$,
with an insertion of a Wilson line operator
\eqn\mci{W_R(C)=\Tr_RP\exp\int_CA.}
Here $R$ is a representation of $G$, and $C\subset \S^1\times D$ 
is a circle of the form
$C=\S^1\times P$, with $P$ being a point in the disc $D$.
For any given $R$, the path integral on $\S^1\times D$ with insertion
of $W_R(C)$  gives a state $\Psi_R$ in the Chern-Simons Hilbert
space on $\S^1\times \S^1$ at level $k'$, and this space, as we have argued,
is the same as ${\cal H}$.
As $R$ ranges over the highest weights of integrable representations
of the $\widehat G$ affine algebra at level $k'$, the $\Psi_R$ furnish 
a basis of ${\cal H}$ \wittenjones.  (The $\Psi_R$'s for other representations
are zero or a multiple of one of the $\Psi_R$'s for an integrable 
representation.) 

Going back to the electric flux operator $T_\omega$,
its action on the state $\Psi_R$ is now clear.
It maps $W_R(C)$ to $\omega(R)W_R(C)$, where the central element
$\omega$ of $G$ acts in the irreducible representation $R$ as
multiplication by $\omega(R)$.   So it likewise maps $\Psi_R$ to
$\omega(R)\Psi_R$.

If there is a zero energy state carrying electric flux for any value
of the spatial volume, this means
that the theory is not confining.  Confinement of electric flux
can therefore occur only if the center of $G$ acts trivially on all
$\Psi_R$, or equivalently on
all integrable representations of the WZW model
at level $k'$.  This, however,
is so only at $k'=0$ (where only the trivial representation is
integrable).  For example, for $SU(2)$, at level $k'$, the integrable
representations have highest weights $0,1/2,1,\dots,k'/2$,
so whenever $k'>0$, there is an integrable representation of
half-integer spin, on which the center acts nontrivially.
Among the theories with unbroken supersymmetry,
only the theory with $k'=0$ -- and thus $k=\pm h/2$ -- might
be interpreted as confining.

The theories with $|k|<h/2$ may very well also be confining,
but as they conjecturally have spontaneously broken supersymmetry,
we cannot probe their dynamics by looking for supersymmetric states.

\newsec{Classification Of Massive Phases}

\nref\xno{T. Filk, M. Marcu, and K.
Fredenhagen, ``Line Of Second-Order Phase Transitions
In The Four-Dimensional $\Z_2$ Gauge Theory With Matter Fields,''
Phys. Lett. {\bf 169B} (1986) 405; M. Marcu, ``(Uses Of) An Order
Parameter For Lattice Gauge Theories With Matter Fields,'' 
in {\it Wuppertal 1985, Proceedings, Lattice Gauge Theory}, p. 267.}

This concluding section will be devoted to some remarks 
 about the classification of massive phases of gauge theories.\foot{These
remarks were suggested in part by discussions ca. 1990 with M. Marcu
-- see \xno\ -- and M. F. Atiyah.}

Consider a gauge theory with a mass gap. Let us look at the behavior
of Wilson loop operators $W_R(C)=
\Tr_R P\exp\int_CA$, with $R$ some representation
of $G$ and $C$ a loop in spacetime.  Let $L(C)$ be the circumference
of $C$, and $A(C)$ the minimal area of a surface that it spans.
The renormalization of $W_R(C)$ that we allow is local along the
loop: 
\eqn\xonxonxp{W_R(C)\to e^{\alpha_R L(C)}W_R(C).}
Here $\alpha_R$ is a renormalization parameter.  
  We want to study the behavior
of $\langle W_R(C)\rangle$ as the loop $C$ is scaled up in size.
In some theories (and for some representations), 
$\langle W_R(C)\rangle \sim \exp(-\beta_R A(C))$, with $\beta_R$ a
positive constant, modulo subleading
terms which vary with the circumference rather than the area.  Such
a statement is invariant under a renormalization of the form \xonxonxp.
If $\langle W_R(C)\rangle$ behaves this way, then it vanishes in the
limit that $C$ is scaled up no matter what renormalization is used.  
On the other hand, it may happen
that the area coefficient $\beta_R$ vanishes.  In this case,
$\langle W_R(C)\rangle$ has with generic renormalization an exponential
dependence on the circumference.  We can pick the renormalization
constant in \xonxonxp\ to cancel this term.   What happens then?

In a theory with a mass gap, one would expect
that for large loops, \eqn\klop{\ln \langle W_R(C)\rangle = \beta_RA(C)
+\gamma_RL(C) +\dots,} where the $\dots$ terms are {\it constant}
in the limit of large loops.  (In the absence of a mass gap, there
can be much more complicated behavior; for example, Feynman diagrams
in a theory with massless fields
can give terms $L(C) \ln L(C)^m$ for all integers $m$.)
If $\beta_R$ vanishes and we choose the renormalization to cancel
$\gamma_R$, then $\ln \langle W_R(C)\rangle$ should have a limit
as $C$ becomes large.
We conclude then that with our renormalization
\eqn\uxxu{N_R(C) = \lim_{C\to\infty}\langle W_R(C)\rangle }
should exist in a massive gauge theory.  The confining case -- $\beta_R>0$
-- is the case that $N_R(C)=0$.

Above three dimensions, this construction exhibits one constant for
every representation.  In three dimensions, the construction is much
richer, because the loop $C$ in spacetime may be {\it knotted}.
Thus, in the three-dimensional case, in a massive gauge theory,
we get an invariant for every representation and every knot class.
In the examples we have been examining in the present paper,
there is a mass gap for sufficiently large $|k|$  (conjecturally
that this is so precisely if $|k|\geq h/2$), and the ``low energy''
theory is a Chern-Simons theory at level $k'=k-h/2\cdot {\rm sgn}\,k$.  
The large
loop limits of the expectation values of the $\langle W_R(C)\rangle$
can \wittenjones\ be expressed in terms of the celebrated Jones polynomial of 
knots and its generalizations.  

At least in the three-dimensional examples, it seems fairly clear
that the $N_R(C)$  must depend only on the universality
class of the theory.  The effective theory at very long distances is
characterized just by the integer $k'$ -- which controls the knot invariants --
and this integer cannot change under continuous variation of parameters.

In four-dimensional massive gauge theories, the meaning and significance
of the  $N_R(C)$
are less apparent.  It seems probable, however, that they are invariants
of the universality class of a theory.  

\bigskip\noindent{\it Analysis Of Phases In Three Dimensions}

The dependence on $k'$ makes
 it clear that the usual Higgs/confinement dichotomy is not
the whole story for classification of massive phases of gauge theories
in three dimensions.
We have, on the contrary, infinitely many inequivalent universality classes,
parameterized by $k'$; none of the theories with $k'>0$ is confining
as they all have some nonzero $N_R(C)$ with $R$ a representation
in which the center of the gauge group acts nontrivially.
This follows from the formulas in \wittenjones\ for expectation
values of Wilson loops in pure Chern-Simons theory.  Alternatively,
the theories with $k'>0$ are not confining, since we showed in section 
3.3 that in these theories, electric
flux winding on a torus has no cost in energy.

The case that $k'=0$, when the low
energy theory is a ``pure gauge theory'' without Chern-Simons
interaction,  might be confining.  Some evidence for
this appeared
 in section 3.3, where we saw that in this case, there
is no zero energy state on $\T^2$ with electric flux.  
It follows from this fact that all $N_R(C)$, with the
center of $G$ acting nontrivially on $R$, vanish if $k'=0$.
For one could factorize the evaluation of $N_R(C)$ by ``cutting''
the three-dimensional spacetime on a two-torus $S$ that consists
of all points a distance $\epsilon$ from $C$, for some small $\epsilon$.
Upon scaling $C\to\infty$, one can also take $\epsilon\to\infty$,
and the path integral on the solid torus bounded by $S$ and containing
$C$ gives a state carrying electric flux in the Hilbert space
obtained by quantizing the pure Chern-Simons theory
on $C$.  As the pure Chern-Simons theory
has no physical states on $S$ that have electric flux, the path
integral for $\langle W_R(C)\rangle$ will vanish for $C\to\infty$.

The above reasoning used a possibly risky
 analytic continuation of the Chern-Simons results
(which are usually considered for $k'>0$) to $k'=0$.  
I will now describe somewhat more explicitly how this analytic
continuation works, taking
$G=SU(2)$ as an illustration.
In $SU(2)$ Chern-Simons theory at level $k'$, the loop expectation
value $N_R(C)$, for a non-trivial representation $R$ with highest weight
of spin $j$, vanishes if $j$ is 
congruent to $-1/2$ mod $(k'+2)/2$, but not otherwise for a generic $C$.
\foot{All Chern-Simons observables 
can be expressed in terms of quantities in the WZW model such as the
matrix $S$ that generates the modular transformation $\tau\to -1/\tau$
on the characters.
In a basis of representations of highest weight $j$,
the matrix elements of $S$ for $SU(2)$ at level $k'$ are
$S_{jj'}=\sqrt {2/(k+2)}\sin\left(\pi(2j+1)(2j'+1)/(k'+2)\right)$.
This shows the vanishing if $j$ or $j'$ is congruent to $-1/2$ mod
$(k'+2)/2$.}
For $k'=0$, this means that 
$N_R(C)$ 
vanishes if $j$ is a half-integer, and  not otherwise. 
This is the usual
statement of confinement: there is an area law precisely
if the representation
transforms nontrivially under the center of the gauge group. 

Thus it seems likely that precisely at $k=\pm h/2$, the theories studied
in the present paper have a mass gap,
unbroken supersymmetry and  confinement.  
For $|k|>h/2$, they are in inequivalent Higgs-like phases with a mass gap
and unbroken supersymmetry,
and for $|k|<h/2$, they conjecturally have a massless Goldstone fermion
(and perhaps confinement).

\def\O{{\cal O}}
Going back to the three-dimensional examples, let us examine
the other standard criterion for confinement, which is whether external
magnetic flux is screened.
In three dimensions, one 
considers a local 't Hooft operator $\O(P;w)$ defined by removing a point $P$
 from spacetime and inserting a nontrivial magnetic flux $w$ on a small
sphere surrounding $P$.  (In four dimensions, one has instead an
't Hooft loop operator defined by removing a loop $C$ from spacetime
and inserting magnetic flux on a sphere that links $C$.)
In the three-dimensional case, a restriction on $k$ is needed
in introducing the operators $\O(P;w)$.  For instance, as we saw in
section 2, if $G=SU(n)$ and $w$ is prime to $n$, then the restriction
is that $k$ should be congruent to $n/2$ modulo $n$.

't Hooft's criterion for a Higgs phase of a massive gauge theory in four
dimensions
is that the 't Hooft loop should show area law in four dimensions;
in three dimensions the criterion is that
 the expectation value $\langle \O(P;w)\rangle$ should vanish.
In our three-dimensional examples, one might expect this criterion
to be obeyed for $k>h/2$, as these theories are not confining.
This is so.  On $\S^2$, the modulo space of flat connections
on a bundle with nonzero magnetic flux is empty, and hence the
Chern-Simons theory if quantized with such a bundle
on $\S^2\times \R$ (with $\R$ understood
as the ``time'' direction) has no physical states.  Because of the topological
invariance of the Chern-Simons theory at long distances, the 
expectation value $\langle \O(P;w)\rangle$ 
can be computed in radial quantization
-- where the radius measures the distance from $P$.  In other words,
we consider
the operator $\O(P;w)$ to prepare an initial state at $r=0$ ($r$ being
the distance from $P$), and propagate outward to $r=\infty$.
This propagation should
project onto zero energy states. 
But there are no zero energy states to project onto, so the expectation
value vanishes.  Or more
prosaically, the expectation value $\langle \O(P;w)\rangle$ vanishes
because -- with a gauge bundle that is nontrivial when restricted
to any arbitrarily large sphere surrounding $P$ -- the classical
equation of motion $F=0$ of the long distance effective Chern-Simons
theory cannot be obeyed even near spatial infinity.  

Now let us consider a different 
but also standard criterion for ``screening of magnetic flux.''
  In this alternative formulation, we quantize
the theory on $\T^2\times \R$ and interpret magnetic screening to mean
that in the limit of large volume of $\T^2$, the ground state
energy is independent of the magnetic flux on $\T^2$.  
With this criterion, magnetic screening does occur in the three-dimensional
$N=1$ gauge theories for all allowed $k\geq h/2$ 
since, as we have seen in sections 2 and 3, $\Tr\,(-1)^F$ is nonzero
and hence the ground state energy vanishes
whether there is magnetic flux or not.  

Thus, the two standard
criteria for magnetic confinement
give different answers in these theories.  The key difference between
the two criteria is that 
there are flat connections on a bundle over $\T^2$ with magnetic
flux, but not on such a bundle over $\S^2$. 
As far as I know, the distinction between the two notions of magnetic
screening has not been important in massive phases of gauge theories
that have been studied previously.

\bigskip\noindent{\it Generalization}

Part of the above story is special to three dimensions, but part is not.

One basic question is how to describe the long distance limit of a theory.
It is conventionally claimed that the long distance limit of a massive
theory is ``trivial,'' but the very idea of 't Hooft and Wilson loops
as criteria for confinement shows that there is more to say about the long
distance limit of a massive theory than just this.

The lesson from the above discussion is that a massive theory may
give at long distances a nontrivial topological field theory, which
governs the possible vacuum states in different conditions,
even though there are no ``physical excitations'' at very long wavelengths.
Topological field theory is particularly interesting in three-dimensions
because of the existence of the Chern-Simons theories.   In four
dimensions, 
the known unitary examples are less interesting.  
The most obvious example of a topological field theory in four dimensions
(or indeed in any dimension)
is a gauge theory with a finite gauge group $\Gamma$.\foot{Donaldson
theory is based on a non-unitary topological field theory in four dimensions.}   
  If such a theory is
quantized on a three-manifold $X$, the number of physical states 
is the number of conjugacy classes of 
representations of the fundamental group of
$X$ into $\Gamma$.  (These theories   have
been discussed in \ref\diwit{R. Dijkgraaf and E. Witten, ``Topological
Field Theories And Group Cohomology,''  Commun. Math. Phys. {\bf 129}
(1990) 393.}, including also a
generalization involving a group cohomology class.  The generalization leads
to a somewhat 
more elaborate formula for the number of physical states.)
In particular, these theories do depend on $\Gamma$.

Consider a weakly coupled four-dimensional theory with mass gap
in which, perturbatively, a connected simple gauge group
$G$ is spontaneously broken to a finite subgroup $\Gamma$.
From the point of view of the usual criteria involving 
't Hooft and Wilson loops, these theories are all Higgs theories. 
For example, by virtue of
 the perturbative Higgs mechanism, the Wilson loops can be computed
 reliably in perturbation theory and
show no area law.  't Hooft loops do show an area law, since a
bundle on $\S^2$ with nontrivial magnetic flux cannot have
a $\Gamma$-valued connection, so a path integral with
such a bundle (on an $\S^2$ that links an 't Hooft loop)
 cannot receive a contribution
in the low energy theory.

Nevertheless, order parameters distinguishing such theories have
been constructed \ref\presk{J. Preskill and L. M. Krauss,
``Local Discrete Symmetry And Quantum Mechanical Hair,''
Nucl. Phys. {\bf B341} (1990) 50.}.  We can reformulate this
discussion to some extent and say that a basic order parameter
is the topological field theory that
prevails at long distances.  It is simply a gauge theory of the finite
group $\Gamma$.

Here is a simple yet interesting example.  Take $G=SO(3)$ and let
$\Gamma$ be the subgroup -- isomorphic to $\Z_2\times \Z_2$ -- consisting
of diagonal matrices with entries $\pm 1$ (and determinant 1).
As we have already explained, magnetic flux wrapped on $\S^2$ is unscreened,
and the 't Hooft loops show area law.  However, in such a theory, magnetic
flux on $\T^2$ is screened.  This is so simply because a bundle
on $\T^2$ with nonzero magnetic flux admits a flat connection with
holonomies in $\Gamma$.  The holonomies around the two directions in
$\T^2$ can be the matrices
\eqn\xkkm{U=\left(\matrix{ 1 & 0 & 0 \cr 0 & -1 & 0 \cr 0 & 0 & -1\cr}\right),
~~V=\left(\matrix{-1& 0 & 0 \cr 0 & -1 & 0 \cr 0 & 0 & 1\cr}\right).}
A flat bundle with these holonomies has nonzero magnetic flux since
the matrices $U$ and $V$, if lifted to $SU(2)$, do not commute.
This gives an elementary four-dimensional
example in which standard criteria for magnetic screening give different
results, somewhat as we found in three dimensions.

\bigskip
I would like to thank Z. Guralnik, A. Kapustin, V. Periwal,
 N. Seiberg, and M. Strassler for discussions.
This work was supported in part by NSF Grant PHY-9513835.
\listrefs
\end